\documentclass[usenatbib,usegraphicx]{mn2e}
\usepackage{subfig}
\usepackage[dvipdfm]{hyperref}
\usepackage{color}
\usepackage{amssymb}

\newcommand {\eg} {{\it e.g.}}

\newcommand {\be} {\begin{equation}}
\newcommand {\ee} {\end{equation}}
\newcommand {\bea} {\begin{eqnarray}}
\newcommand {\eea} {\end{eqnarray}}

\title[The brightest blazar flares]{The brightest gamma-ray flares of blazars}

\author[Nalewajko]{Krzysztof~Nalewajko\thanks{\tt{knalew@colorado.edu}}\\
University of Colorado, UCB 440, Boulder, CO 80309, USA}

\begin{document}

\maketitle

\begin{abstract}
I present a systematic study of gamma-ray flares in blazars. For this purpose, I propose a very simple and practical definition of a flare as a period of time, associated with a given flux peak, during which the flux is above half of the peak flux. I select a sample of 40 brightest gamma-ray flares observed by \emph{Fermi}/LAT during the first 4 years of its mission. The sample is dominated by 4 blazars: 3C~454.3, PKS~1510-089, PKS~1222+216 and 3C~273. For each flare, I calculate a light curve and variations of the photon index. For the whole sample, I study the distributions of the peak flux, peak luminosity, duration, time asymmetry, average photon index and photon index scatter. I find that: 1) flares produced by 3C~454.3 are longer and have more complex light curves than those produced by other blazars; 2) flares shorter than 1.5 days in the source frame tend to be time-asymmetric with the flux peak preceding the flare midpoint. These differences can be largely attributed to a smaller viewing angle of 3C~454.3 as compared to other blazars. Intrinsically, the gamma-ray emitting regions in blazar jets may be structured and consist of several domains. I find no regularity in the spectral gamma-ray variations of flaring blazars.
\end{abstract}


\section{Introduction}

Blazars constitute the most numerous class of permanent gamma-ray sources in the sky. In the 2-year \emph{Fermi} Large Area Telescope (LAT) Source Catalog \citep[2FGL;][]{2012ApJS..199...31N}, they constitute 67\% of \emph{associated} point sources. They are characterised by strong and chaotic variability, occasionally producing spectacular flares. These events have attracted a considerable attention, triggering many extensive multiwavelength campaigns that attempted to characterise the overall behaviour of individual objects \citep[\eg,][]{2010ApJ...712..405V,2010ApJ...721.1425A,2011ApJ...726L..13A,2012ApJ...751..159A,2012ApJ...754..114H}. Unfortunately, the results of these campaigns vary widely, revealing relatively few consistent patterns that can be used to constrain the underlying physics of relativistic jets.

The \emph{Fermi} Space Telescope offers a unique capability in the observational astronomy. Its main detector, LAT (\citealt{2009ApJ...697.1071A}), has a very wide field of view, of $\sim 50^\circ$. For most of its scientific mission starting in August 2008, \emph{Fermi} operated in the scanning mode, allowing a uniform coverage of the whole sky every several hours. Therefore, almost every blazar flare that happened during the \emph{Fermi} mission was observed from the beginning to the end with basically the same sensitivity. It is then straightforward to produce a fairly complete and unbiased sample of gamma-ray flares of blazars, and to compare their properties within individual sources and between them. Such an approach has a potential to clearly demonstrate the similarities and differences among individual blazar flares.

In this work, I study a sample of 40 bright gamma-ray flares of blazars detected by \emph{Fermi}/LAT during the first 4 years of its mission. In Section \ref{sec_sel}, I propose a useful definition of a gamma-ray flare, and describe the process of selecting the sample. In Section \ref{sec_res}, I present light curves of these flares, and study the distributions of their duration, time asymmetry, average photon index and photon index scatter. In Section \ref{sec_dis}, I discuss the general properties of the flares and their interpretation. I conclude in Section \ref{sec_con}.

\begin{table*}
\centering
\caption{List of the brightest gamma-ray flares of blazars. ${\rm MJD}_{\rm peak}$ is the moment of flux peak, $F_{\rm peak}$ is the peak flux in units of $10^{-6}\;{\rm ph\,s^{-1}\,cm^{-2}}$, $t_1$ is the flux doubling time scale, $t_2$ is the flux halving time scale, $T=(t_1+t_2)$ is the flare duration, $k=(t_1-t_2)/(t_1+t_2)$ is the time asymmetry parameter, $\left<\Gamma\right>$ is the average photon index, and ${\rm rms}(\Gamma)$ is the photon index scatter. All times are in units of days.
References:
a -- \citet{2011ApJ...733L..26A},
b -- \citet{2010ApJ...721.1383A},
c -- \citet{2011ApJ...733...19T},
d -- \citet{2010ApJ...721.1425A},
e -- \citet{2010ApJ...714L..73A}.}
\label{tab_top40}
\begin{tabular}{rllrrrrrrrr}
\hline\hline
\# & blazar & ${\rm MJD}_{\rm peak}$ & $F_{\rm peak}$ & $t_1$ & $t_2$ & $T$ & $k$ & $\left<\Gamma\right>$ & ${\rm rms}({\Gamma})$ & ref \\
\hline
1  & 3C~454.3     & 55520.0 & 77.2 $\pm$ 2.4 & 2.9 & 1.0 & 3.9 &  0.49 $\pm$ 0.07 & 2.24 & 0.10 & a \\
2  & 3C~454.3     & 55526.9 & 33.3 $\pm$ 1.8 & 1.1 & 3.9 & 5.0 & -0.58 $\pm$ 0.05 & 2.32 & 0.04 \\
3  & PKS~1510-089 & 55853.8 & 26.6 $\pm$ 2.0 & 0.2 & 0.4 & 0.6 & -0.45 $\pm$ 0.46 & 1.99 & 0.03 \\
4  & 3C~454.3     & 55550.3 & 23.6 $\pm$ 1.3 & 3.2 & 5.1 & 8.3 & -0.23 $\pm$ 0.03 & 2.32 & 0.09 \\
5  & 3C~454.3     & 55167.8 & 21.8 $\pm$ 1.5 & 1.2 & 3.1 & 4.3 & -0.44 $\pm$ 0.06 & 2.32 & 0.07 & b \\
6  & 3C~454.3     & 55567.8 & 19.2 $\pm$ 1.6 & 2.9 & 2.5 & 5.4 &  0.07 $\pm$ 0.05 & 2.29 & 0.08 \\
7  & 3C~454.3     & 55301.5 & 16.4 $\pm$ 1.8 & 1.3 & 2.2 & 3.5 & -0.26 $\pm$ 0.07 & 2.35 & 0.13 \\
8  & PKS~1222+216 & 55316.6 & 15.6 $\pm$ 0.9 & 0.4 & 0.4 & 0.8 &  0.00 $\pm$ 0.31 & 1.89 & 0.03 & c \\
9  & PKS~1510-089 & 55872.8 & 15.3 $\pm$ 1.2 & 0.2 & 0.7 & 0.9 & -0.56 $\pm$ 0.29 & 2.13 & 0.08 \\
10 & 3C~454.3     & 55294.1 & 15.3 $\pm$ 0.8 & 5.9 & 3.6 & 9.5 &  0.24 $\pm$ 0.03 & 2.37 & 0.10 & b \\
11 & PKS~1510-089 & 55867.8 & 14.2 $\pm$ 1.5 & 0.5 & 1.5 & 2.0 & -0.50 $\pm$ 0.13 & 2.21 & 0.13 \\
12 & PKS~1222+216 & 55365.8 & 14.2 $\pm$ 1.0 & 1.0 & 1.3 & 2.3 & -0.16 $\pm$ 0.11 & 2.07 & 0.05 & c \\
13 & 3C~454.3     & 55163.1 & 11.8 $\pm$ 1.4 & 2.0 & 1.0 & 3.0 &  0.33 $\pm$ 0.08 & 2.34 & 0.09 \\
14 & 3C~454.3     & 55305.5 & 11.1 $\pm$ 1.5 & 0.8 & 1.1 & 1.9 & -0.16 $\pm$ 0.13 & 2.38 & 0.12 \\
15 & PKS~1510-089 & 55746.2 & 11.1 $\pm$ 1.4 & 0.4 & 0.5 & 0.9 & -0.11 $\pm$ 0.27 & 2.30 & 0.12 \\
16 & PKS~1510-089 & 54948.0 & 10.6 $\pm$ 2.3 & 0.9 & 1.3 & 2.2 & -0.18 $\pm$ 0.11 & 2.38 & 0.08 & d \\
17 & 3C~273       & 55095.3 & 10.3 $\pm$ 0.9 & 0.4 & 1.6 & 2.0 & -0.60 $\pm$ 0.13 & 2.42 & 0.10 & e \\
18 & 3C~273       & 55090.5 & 10.2 $\pm$ 0.9 & 0.4 & 1.2 & 1.6 & -0.50 $\pm$ 0.16 & 2.37 & 0.12 & e \\
19 & PKS~1510-089 & 55980.6 & 10.0 $\pm$ 1.3 & 0.5 & 0.8 & 1.3 & -0.23 $\pm$ 0.19 & 2.24 & 0.07 \\
20 & PKS~1510-089 & 54917.0 &  9.9 $\pm$ 0.7 & 1.9 & 0.5 & 2.4 &  0.58 $\pm$ 0.11 & 2.20 & 0.09 & d \\
21 & PKS~1510-089 & 55851.9 &  9.9 $\pm$ 1.0 & 0.2 & 0.5 & 0.7 & -0.38 $\pm$ 0.39 & 2.32 & 0.13 \\
22 & 3C~454.3     & 55195.2 &  9.7 $\pm$ 0.8 & 0.6 & 1.6 & 2.2 & -0.45 $\pm$ 0.12 & 2.28 & 0.06 & b \\
23 & PKS~1222+216 & 55310.7 &  9.6 $\pm$ 0.8 & 0.5 & 0.7 & 1.2 & -0.22 $\pm$ 0.21 & 1.98 & 0.08 & c \\
24 & 3C~454.3     & 55323.5 &  9.1 $\pm$ 0.9 & 4.2 & 1.8 & 6.0 &  0.40 $\pm$ 0.04 & 2.38 & 0.10 \\
25 & 3C~454.3     & 55327.2 &  8.8 $\pm$ 1.2 & 1.9 & 1.2 & 3.1 &  0.23 $\pm$ 0.08 & 2.33 & 0.08 \\
26 & 3C~273       & 55202.9 &  8.7 $\pm$ 1.1 & 0.3 & 0.6 & 0.9 & -0.33 $\pm$ 0.28 & 2.45 & 0.09 \\
27 & PKS~1222+216 & 55342.1 &  8.7 $\pm$ 0.8 & 0.8 & 0.8 & 1.6 & -0.03 $\pm$ 0.16 & 1.93 & 0.10 \\
28 & PKS~1510-089 & 55990.8 &  8.5 $\pm$ 0.7 & 5.2 & 1.1 & 6.3 &  0.65 $\pm$ 0.04 & 2.29 & 0.08 \\
29 & PKS~1510-089 & 55983.1 &  8.4 $\pm$ 0.8 & 0.5 & 0.4 & 0.9 &  0.11 $\pm$ 0.27 & 2.19 & 0.04 \\
30 & PKS~1510-089 & 54961.8 &  8.2 $\pm$ 0.7 & 0.3 & 0.3 & 0.6 &  0.00 $\pm$ 0.41 & 2.59 & 0.12 & d \\
31 & 3C~454.3     & 55214.3 &  8.0 $\pm$ 0.9 & 0.8 & 0.8 & 1.6 &  0.03 $\pm$ 0.16 & 2.42 & 0.05 \\
32 & PKS~1510-089 & 56002.4 &  7.8 $\pm$ 1.0 & 1.8 & 0.9 & 2.7 &  0.32 $\pm$ 0.09 & 2.45 & 0.11 \\
33 & 3C~454.3     & 55154.8 &  7.8 $\pm$ 0.9 & 0.5 & 1.9 & 2.4 & -0.58 $\pm$ 0.11 & 2.35 & 0.13 \\
34 & PKS~1510-089 & 55767.6 &  7.6 $\pm$ 0.8 & 0.4 & 0.5 & 0.9 & -0.11 $\pm$ 0.27 & 2.09 & 0.07 \\
35 & PKS~1222+216 & 55377.5 &  7.6 $\pm$ 0.7 & 0.4 & 1.1 & 1.5 & -0.45 $\pm$ 0.17 & 2.18 & 0.06 \\
36 & 3C~454.3     & 55283.9 &  7.6 $\pm$ 1.1 & 1.3 & 1.4 & 2.7 & -0.04 $\pm$ 0.09 & 2.37 & 0.13 \\
37 & PKS~1510-089 & 55876.1 &  7.4 $\pm$ 0.8 & 0.7 & 1.0 & 1.7 & -0.18 $\pm$ 0.14 & 2.29 & 0.07 \\
38 & PKS~1222+216 & 55234.0 &  7.4 $\pm$ 0.8 & 0.7 & 0.4 & 1.1 &  0.27 $\pm$ 0.23 & 2.25 & 0.02 \\
39 & PKS~0402-362 & 55827.5 &  7.3 $\pm$ 1.0 & 0.3 & 1.8 & 2.1 & -0.71 $\pm$ 0.13 & 2.27 & 0.11 \\
40 & 3C~454.3     & 55091.6 &  7.1 $\pm$ 0.8 & 4.4 & 1.1 & 5.5 &  0.60 $\pm$ 0.05 & 2.42 & 0.13 \\
\hline\hline
\end{tabular}
\end{table*}

\begin{figure*}
\includegraphics[width=\textwidth]{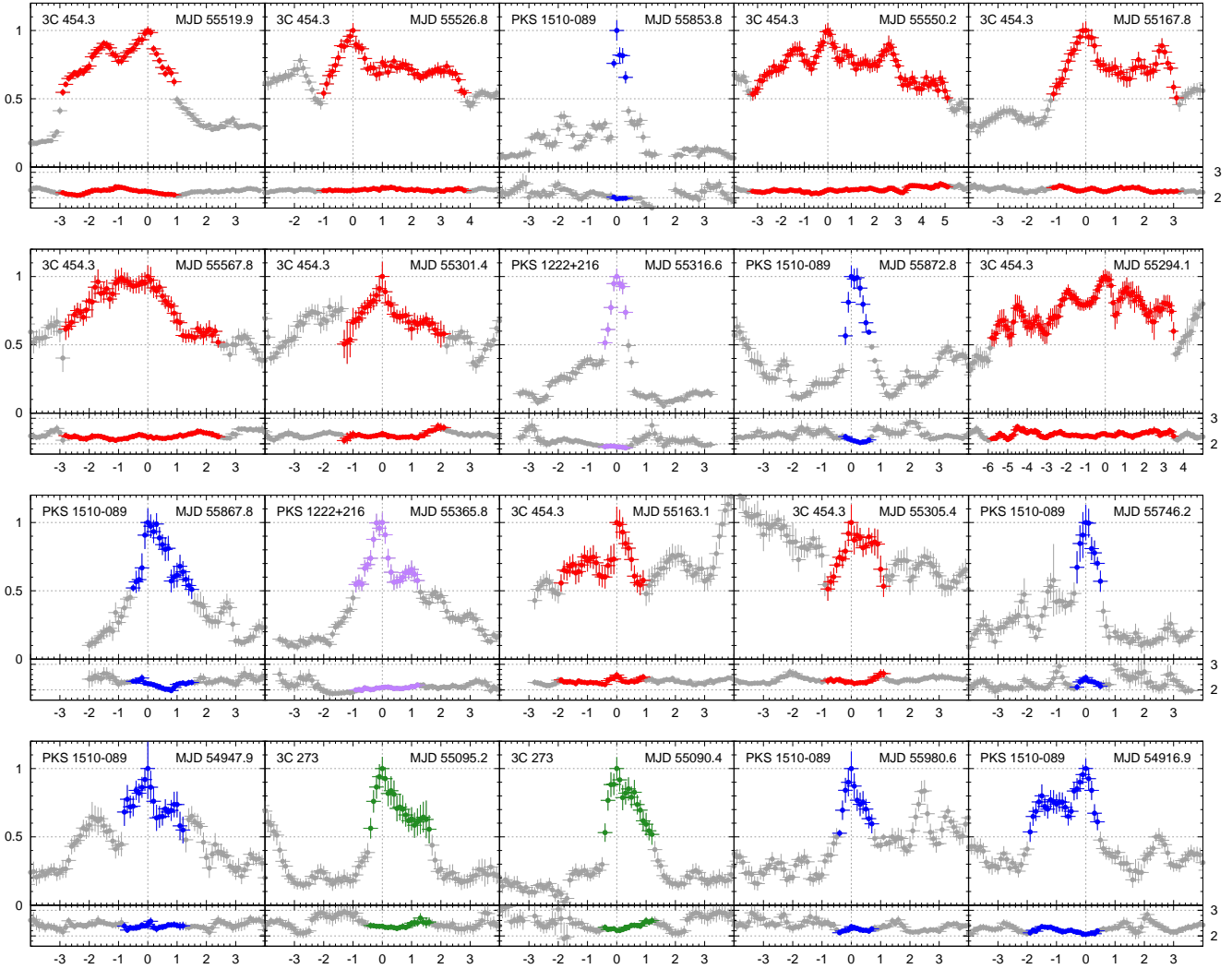}
\caption{Light curves of individual gamma-ray flares calculated from the \emph{Fermi}/LAT data, presented in the order of descending peak photon flux density. For each flare, the upper panel shows the normalised photon flux density integrated in the energy range $0.1-300\;{\rm GeV}$ in 0.5-day time intervals with a 0.1-day step, and the lower panel shows the corresponding photon index. The horizontal axes show time in days measured relative to the flare peak. Data points belonging to the flare period are highlighted with colours.}
\label{fig_lc_top20}
\end{figure*}
\begin{figure*}
\ContinuedFloat
\includegraphics[width=\textwidth]{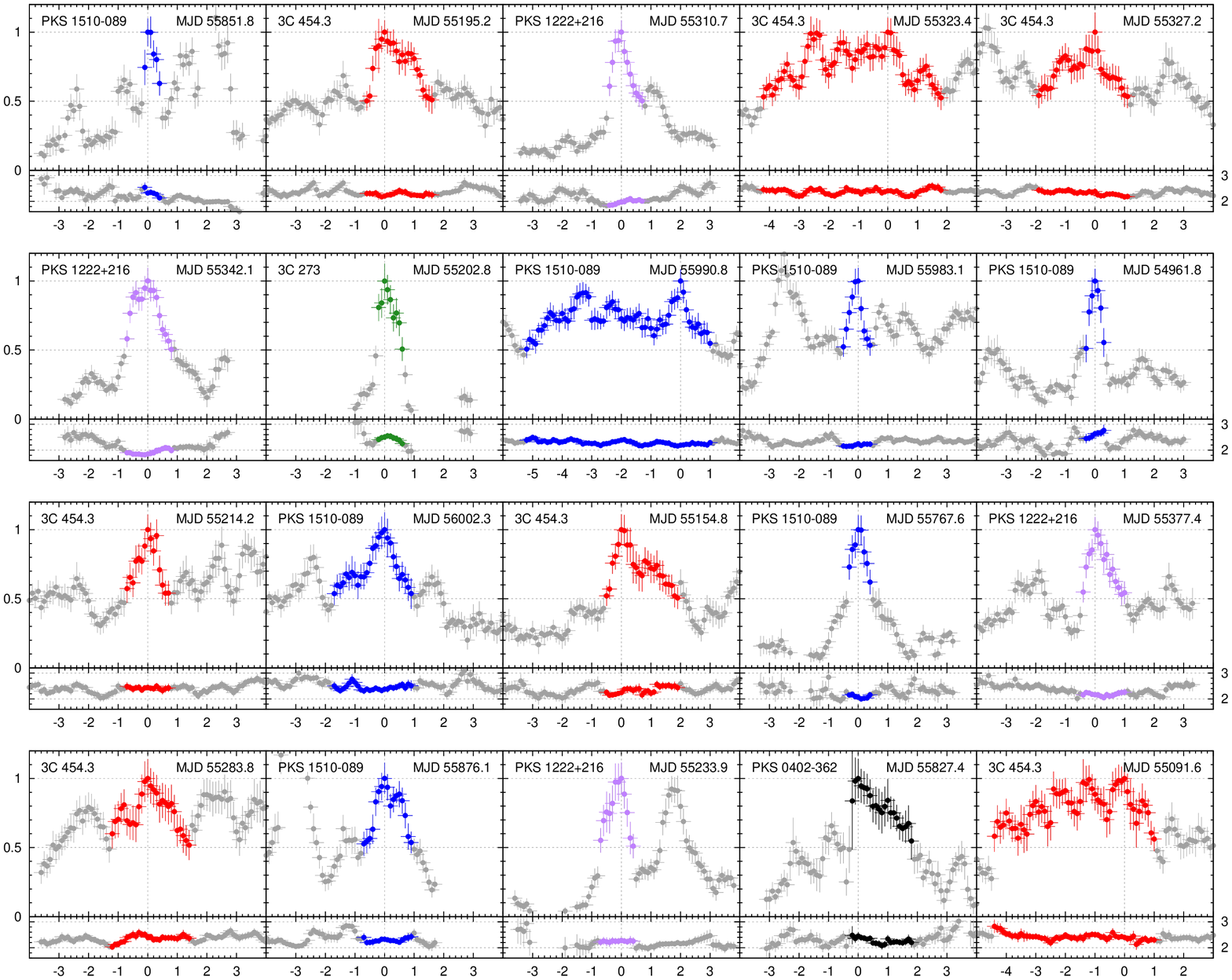}
\caption{(continued)}
\label{fig_lc_top21-40}
\end{figure*}

\begin{figure*}
\includegraphics[width=0.495\textwidth]{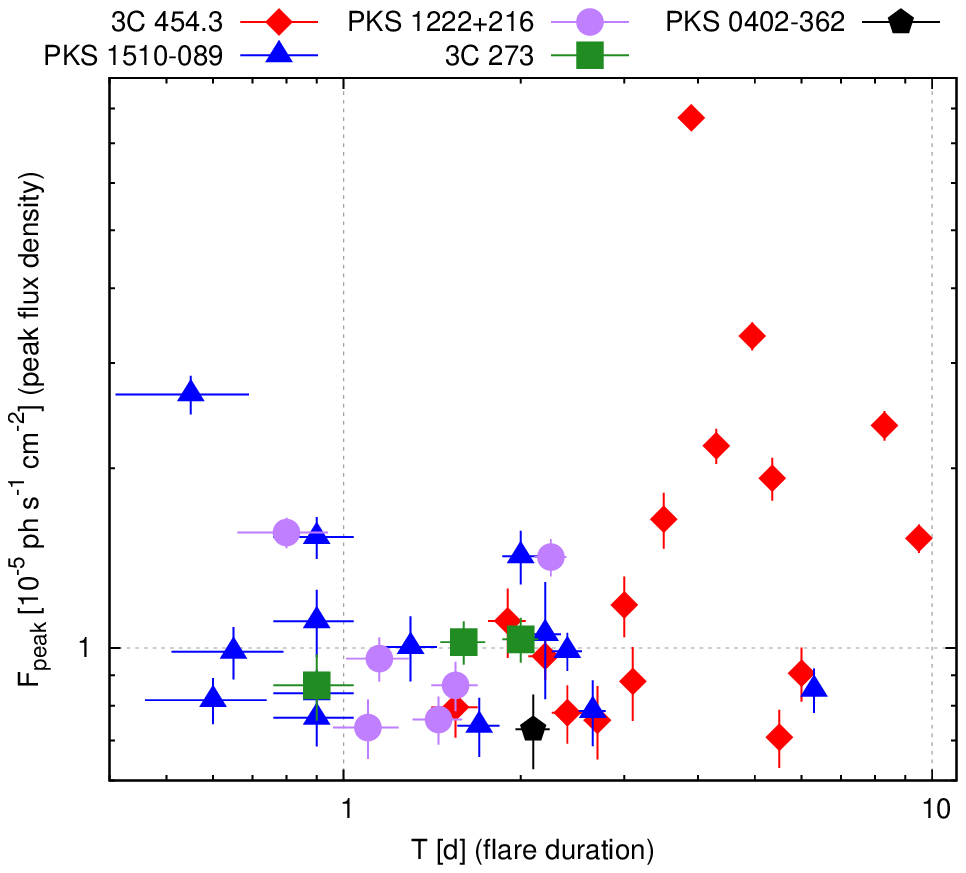}
\includegraphics[width=0.495\textwidth]{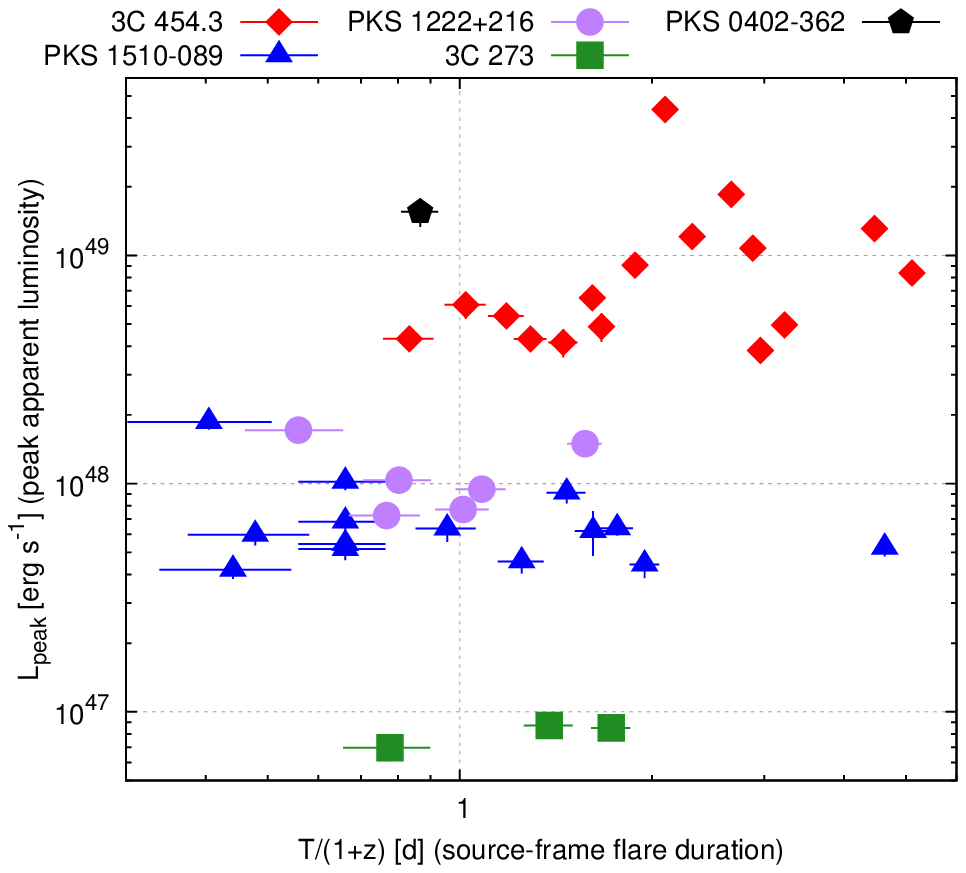}
\caption{\emph{Left panel:} distribution of the observed flare duration $T$ vs. the peak flux density $F_{\rm peak}$. \emph{Right panel:} distribution of the source-frame flare duration $T/(1+z)$ vs. the peak apparent luminosity $L_{\rm peak}$. The shape and colour of each point indicates the host blazar.}
\label{fig_scat_flux_dur}
\end{figure*}

\section{Sample selection}
\label{sec_sel}

Any discussion of blazar flares immediately leads to a fundamental difficulty -- how to define a flare. The nature of blazars variability is stochastic \citep{2010ApJ...722..520A}, characterised by a very high duty cycle, and there is no consistent `background' flux level. It is thus difficult to draw a division between `flares' and `quiescence'. In fact, many researchers question the validity of using the term `flare' in the context of blazar light curves. On the other hand, this term is intuitive, as blazars spend very little time in their highest flux states \citep{2010MNRAS.405L..94T}. My opinion is that the term is certainly useful, but for a systematic study a rigorous definition is necessary.

Here, I propose a very simple definition, in which a flare is a contiguous period of time, associated with a given flux peak, during which the flux exceeds half of the peak value, and this lower limit is attained exactly twice -- at the beginning and at the end of the flare. One can show that this definition does not allow any two flares to overlap. Every two flares must be separated by a flux minimum which is lower than half of the lower peak. A flare may include minor peaks, but they are not considered to be peaks of separate subflares. This definition is natural in a stochastic light curve with a power density spectrum (PDS) in the form of a power-law, as it is based on equal flux ranges on a logarithmic scale. It is practical, as it is very easy to determine local flux peaks and the associated lower flux limits. Also, it automatically provides two standard variability time scales often discussed in the blazar studies -- the flux-doubling time scale and the flux-halving time scale. Figure \ref{fig_lc_top20} illustrates this definition in various real light curves.

The process of selecting a flux-limited sample of blazar flares is divided into two steps. In the first step, a preliminary list of blazar \emph{active periods} is extracted from the \emph{Fermi}/LAT Monitored Source List\footnote{\url{http://fermi.gsfc.nasa.gov/ssc/data/access/lat/msl_lc/}}. It is a database of daily and weekly flux estimates for all sources exceeding the $0.1-300\;{\rm GeV}$ photon flux of $10^{-6}\;{\rm ph\,s^{-1}\,cm^{-2}}$. I select all daily photon fluxes during the first 4 years of the \emph{Fermi} mission (MJD 54682 -- 56143) above $3\times 10^{-6}\;{\rm ph\,s^{-1}\,cm^{-2}}$ for sources located away from the Galactic plane ($|b|>10^\circ$). For each blazar, these daily entries are grouped into active periods, which are separated by gaps of at least 10 days. If an active period contains two flux peaks separated by at least 10 days and a flux minimum lower than the half of the lower peak flux or at least one-day gap, it is split into two active periods along the flux minimum. The active periods for all blazars are sorted according to the decreasing peak flux, and the first 30 active periods are selected for the second step. This choice corresponds to a minimum flux peak of $4.5\times 10^{-6}\;{\rm s^{-1}\,cm^{-2}}$, and the preliminary list of active periods consists of 6 blazars: 3C~454.3, PKS~1510-089, PKS~1222+216, 3C~273, PKS~0402-362 and PKS~1329-049.

In the second step, I perform the maximum likelihood analysis of the \emph{Fermi}/LAT data. I use the standard analysis software package {\tt Science Tools v9r27p1}, with the instrument response function {\tt P7SOURCE\_V6}, the Galactic diffuse emission model {\tt gal\_2yearp7v6\_v0}, and the isotropic background model {\tt iso\_p7v6source}. Events of the {\tt SOURCE} class were extracted from regions of interest of $10^\circ$ radius centred on the position of each blazar. The background models included all sources from the 2FGL within $15^\circ$, their spectral models are power-laws with the photon index fixed to the catalogue values. I checked the raw count maps for any new source not included in the 2FGL. In the case of PKS~1510-089, the source model included TXS~1530-131 \citep{2011ATel..3579}. I selected events in the energy range between $100\;{\rm MeV}$ and $300\;{\rm GeV}$. The spectra of all studied blazars were modelled with power-laws. I used overlapping time bins, with the shift between consecutive bins equal to 1/5 of the bin lengths. This allows to avoid the dependency of the results on the `phase' of the time bins and provides a natural interpolation of the light curve. While the consecutive flux measurements are not entirely independent, this is not a problem for the subsequent analysis of the flare parameters.

Light curves calculated with different time bins reveal different amounts of detail, which affects the classification of certain events as flares, and the flares parameters. Shorter time bins reveal more details and potentially shorter flares, but limited photon statistics determines the size of the sample of flares that can be studied reliably. The very brightest gamma-ray flares ($>10^{-5}\;{\rm ph\,s^{-1}\,cm^{-2}}$) can be probed once per orbital period of the \emph{Fermi} satellite ($\sim 1.5\;{\rm h}$; \citealt{2011A&A...530A..77F}). Here, I decide to use the time bin length of $0.5\;{\rm d}$, with the shifts of $0.1\;{\rm d}$, always starting from a full modified julian day (MJD).

Systematic analysis of all candidate active periods revealed that some of them consist of a few flares, so that the list of all flares revealed by this analysis exceeds 50 entries. From this, I select the final list of the 40 brightest flares, which corresponds to the minimum peak flux of $7.1\times 10^{-6}\;{\rm ph\,s^{-1}\,cm^{-2}}$. This limiting peak flux value corresponds roughly to the peak of a histogram of the logarithmic peak flux distribution of all identified flares. The final list of brightest gamma-ray blazar flares is reported in Table \ref{tab_top40}. For each flare, I calculate the flux doubling time $t_1$, the flux halving time $t_2$, and the total flare duration $T=t_1+t_2$. I also calculate the time asymmetry parameter $k=(t_1-t_2)/(t_1+t_2)$, which can take values between $-1$ and $1$, with $k=0$ corresponding to a time-symmetric flare ($t_1=t_2$); the average photon index $\Gamma$ ($N_E\propto E^{-\Gamma}$), and the root mean square (rms) of deviations of the photon index from the mean value. Individual flares will be referred to using their row number in Table \ref{tab_top40}.

Very short flares (e.g. flare \#3 --- PKS~1510$-$089 at MJD 55854) indicate that the measured flux can change significantly over a single time bin shift of $\delta t = 0.1\;{\rm d}$. This provides a practical estimate of accuracy in determining the moments of flare beginning, peak and ending. The uncertainty of parameters $t_1$, $t_2$ and $T$ can be estimated as $\sqrt{2}\delta t = 0.14\;{\rm d}$. The uncertainty of the asymmetry parameter $k$ is given by $\delta k = [2(3+k^2)]^{1/2}(\delta t/T)$, which is large for the shortest flares. The values of $\delta k$ are reported in Table \ref{tab_top40}.

\begin{figure}
\includegraphics[width=\columnwidth]{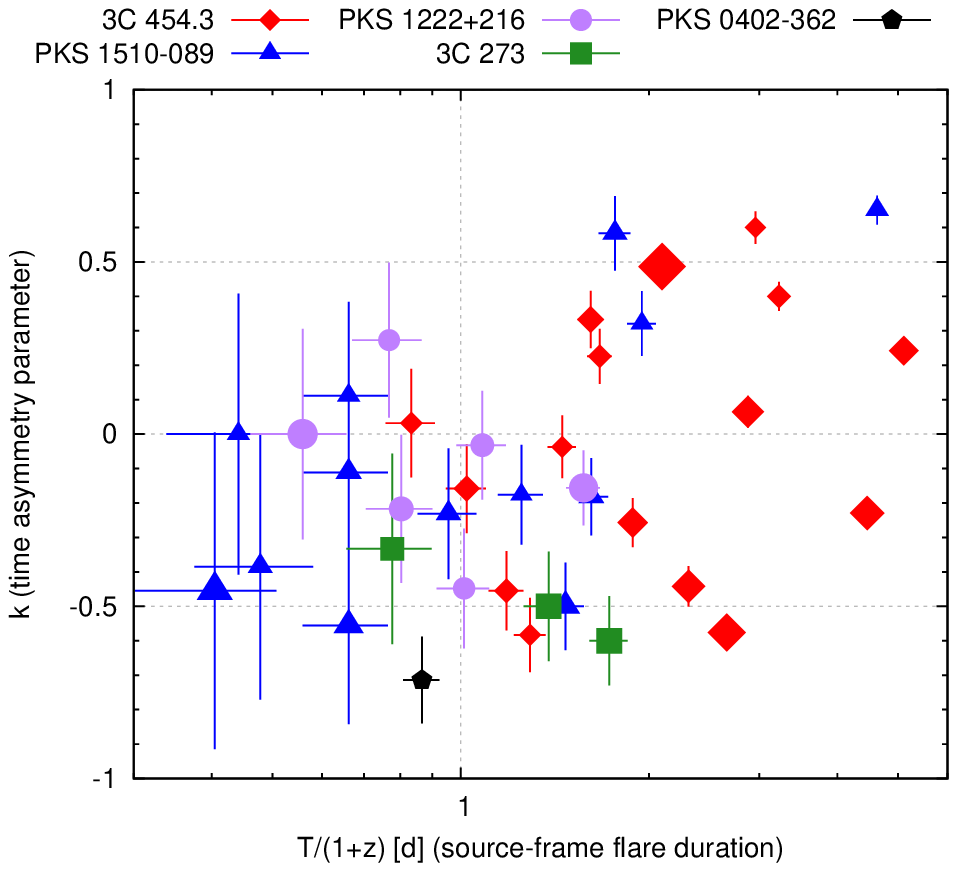}
\caption{Distribution of the source-frame flare duration $T/(1+z)$ vs. the time asymmetry parameter $k$. The shape and colour of each point indicates the host blazar, and the point size is proportional to the logarithm of peak flux $F_{\rm peak}$.}
\label{fig_scat_dur_asy}
\end{figure}

\section{Results}
\label{sec_res}

My sample of the brightest gamma-ray flares of blazars reported in Table \ref{tab_top40} is produced by only 5 blazars: 3C~454.3 (16 entries), PKS~1510$-$089 (14 entries), PKS~1222+216 (6 entries), 3C~273 (3 entries), and PKS~0402$-$362 (1 entry). Remarkably, 3C~454.3 produced 7 of the 10 brightest flares. About 20 flares exceed the peak flux level of $10^{-5}\;{\rm ph\,s^{-1}\,cm^{-2}}$.

In the following analysis, I will compare both the observed parameters, and parameters measured in the source frame. The redshifts and luminosity distances of the 5 blazars that produced the sample are given in Table \ref{tab_blazars}.

\begin{table}
\centering
\caption{Redshifts and luminosity distances of blazars producing the sample of brightest gamma-ray flares. I adopt standard $\Lambda$CDM cosmology with $H_0=71\;{\rm km\,s^{-1}\,Mpc^{-1}}$, $\Omega_{\rm m}=0.27$ and $\Omega_{\rm \Lambda}=0.73$.}
\label{tab_blazars}
\begin{tabular}{lrr}
\hline\hline
blazar & $z$ & $d_{\rm L}\;{\rm [Gpc]}$ \\
\hline
3C~454.3     & 0.859 &  5.49 \\
PKS~1510-089 & 0.360 &  1.92 \\
PKS~1222+216 & 0.432 &  2.37 \\
3C~273       & 0.158 &  0.75 \\
PKS 0402-362 & 1.423 & 10.31 \\
\hline\hline
\end{tabular}
\end{table}

\subsection{Light curves}

Figure \ref{fig_lc_top20} shows normalised light curves and photon index variations of all 40 flares, in the order of decreasing peak flux, the same as in Table \ref{tab_top40}. The data points belonging to the flare period are highlighted with a colour. One can immediately notice a great variety of the flare shapes and durations. Looking only at the first 10 panels, it is striking that the flares produced by 3C~454.3 appear very different from those produced by PKS~1510$-$089 and PKS~1222+216 --- the former are longer and more complex than the latter. However, looking at the next 30 panels, the picture is less clear. Many flares, especially the longest ones, are characterised by a complex structure with multiple peaks separated by shallow minima. The time scale of these substructure is comparable to the time scale of short flares characterise by a single peak. Medium-long flares ($T\sim 2\;{\rm d}$) often show a single substructure following the main peak (e.g., flare \#12 --- PKS~1222+216 at MJD~55366), and sometimes preceding the main peak (e.g., flare \#20 --- PKS~1510$-$089 at MJD 54917). This results in the significant time asymmetry of these flares.

\begin{figure*}
\includegraphics[width=\textwidth]{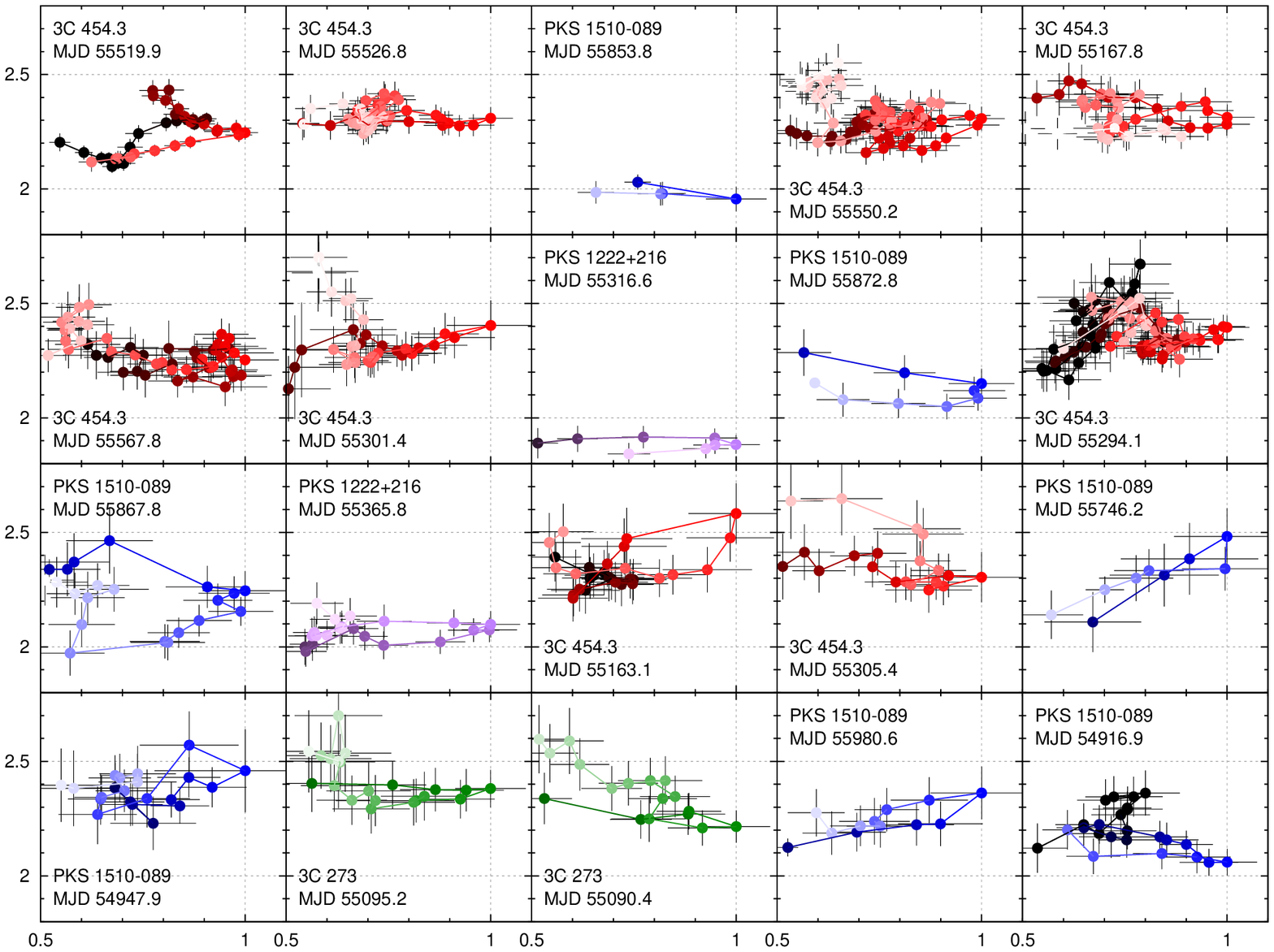}
\caption{Scatter plots showing the normalised $0.1-300\;{\rm GeV}$ photon flux (horizontal axes) vs. the corresponding photon index (vertical axes) for individual flares. The intensity of data points indicates the time flow from dark to bright.}
\label{fig_scat_top20}
\end{figure*}
\begin{figure*}
\ContinuedFloat
\includegraphics[width=\textwidth]{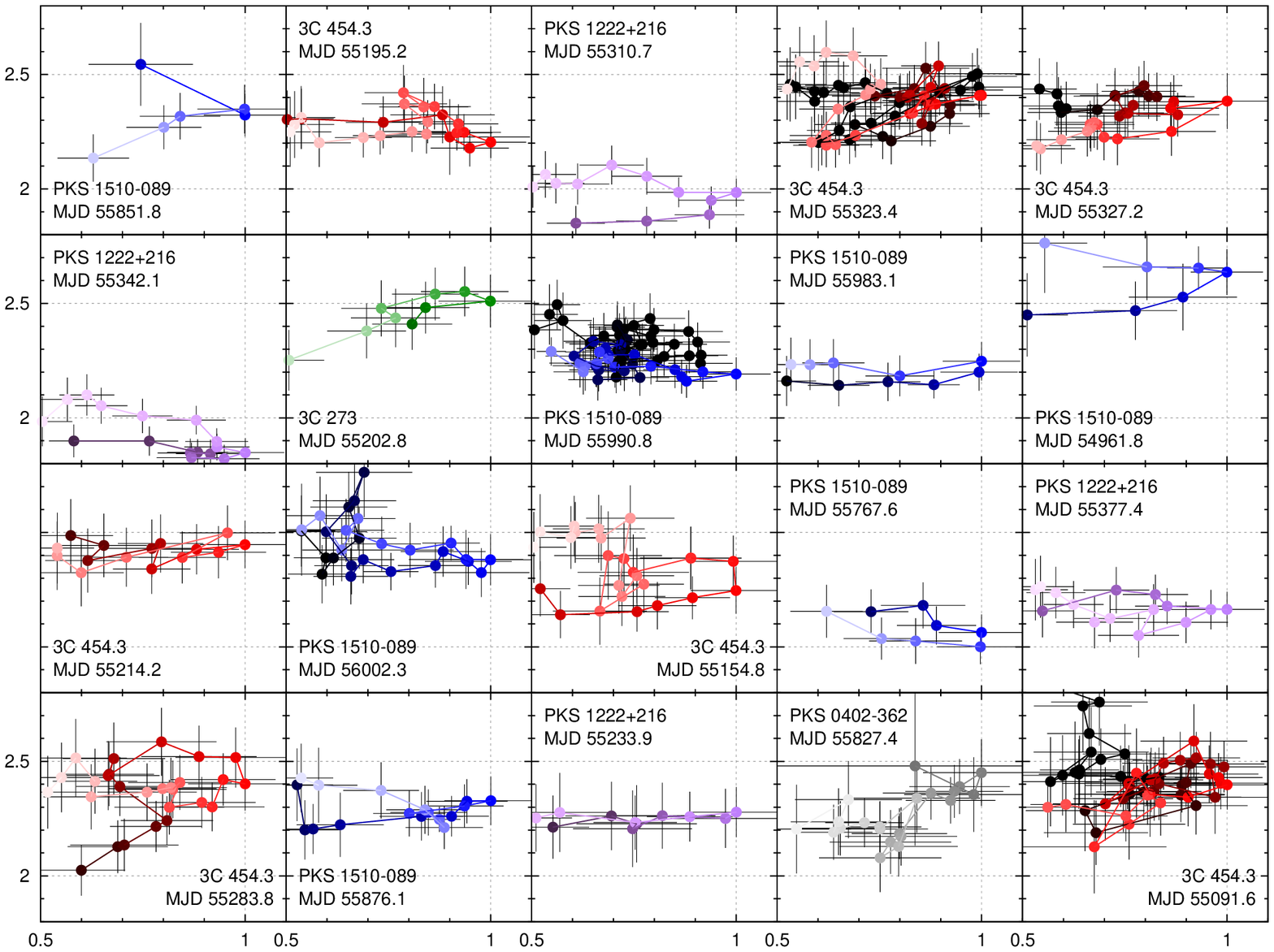}
\caption{(continued)}
\label{fig_scat_top21-40}
\end{figure*}

\subsubsection{Flare duration}

In order to further investigate the temporal properties of blazar flares, in Figure \ref{fig_scat_flux_dur}, I plot the distribution of the flare duration vs. peak flux. Flare durations span the range between 0.5 and 10 days. The lower limit corresponds to the time bin length, and the number of flares with $T<1\;{\rm d}$ (9) is the same as that of flares with $1\;{\rm d}\le T<2\;{\rm d}$. While the sample might be somewhat incomplete for $T<1\;{\rm d}$, it is not likely for longer flares. The average flare duration is $2.7\;{\rm d}$, and the median is $2.1\;{\rm d}$. There appears to be no general trend between the flare duration and the peak flux. However, there is a significant difference between 3C~454.3 on one hand, and PKS~1510$-$089, PKS~1222+216 and 3C~273 on the other. Flares produced by 3C~454.3 are on average longer than those produced by other blazars. 3C~454.3 did not produce a flare shorter than $1.6\;{\rm d}$, while other blazars generally produced flares shorter than $3\;{\rm d}$, with the sole exception of flare \#28 (PKS~1510$-$089 at MJD~55991). For the peak flux greater than $1.2\times 10^{-5}\;{\rm ph\,s^{-1}\,cm^{-2}}$, there is a complete separation of the two groups. Also, there is no discernible difference between the distributions of flares produced by PKS~1510$-$089, PKS~1222+216 and 3C~273.

Also in Figure \ref{fig_scat_flux_dur}, I plot the distribution of the source-frame flare duration vs. peak apparent luminosity. I do not use luminosity integrated from the power-law spectral fit in the $0.1-300\;{\rm GeV}$ range, because: 1) the corresponding source-frame energy range depends on the source redshift, 2) the gamma-ray spectra of these blazars are actually curved/broken, so the power-law spectral fits overestimate the bolometric luminosity above a few GeV. Instead, I use the $\nu L_\nu$ value corresponding to the source-frame photon energy of $500\;{\rm MeV}$. Apparent luminosities of the brightest blazar flares span three orders of magnitude, mainly due to the differences in luminosity distances between the sources. The flare produced by PKS~0402-362, while 39th in observed photon flux, is 3rd in apparent luminosity, due to the high source redshift. There is no clear trend between flare luminosity and duration.

\subsubsection{Flare asymmetry}

Further insight into the flare profiles can be gained by looking at the time asymmetry parameter $k$. In Figure \ref{fig_scat_dur_asy}, I plot the distribution of the source-frame flare duration $T/(1+z)$ vs. $k$. The values of $k$ range between -0.7 and 0.7. This means that the observed flares do not show a very rapid onset directly to the peak or a very rapid decay directly from the peak. The average value of $k$ is $-0.1$, and the median is $-0.16$. Therefore, there is a slight tendency for the flux doubling time $t_1$ to be shorter than the flux halving time $t_2$. Figure \ref{fig_scat_dur_asy} reveals a significant difference between short and long flares. Flares shorter than $\sim 1.5$ days in the source-frame have predominantly negative $k$, while flares longer than $\sim 1.5$ days have a broad and roughly uniform distribution of $k$. Individual blazars produce flares of different distributions of $k$, which can be largely understood as resulting from different distributions of the flare durations. 3C~454.3 produces flares of a broad distribution of $k$, with 4 out of the 5 shortest flares characterised by negative $k$. PKS~1510$-$089 produces flares with mostly negative $k$, however, the 3 longest flares have significantly positive values of $k$. PKS~1222+216 produces fairly symmetric flares, while the 3 flares produced by 3C~273 have significantly negative $k$. The most asymmetric flare in the sample is the one produced by PKS~0402$-$362. There is a slight trend for the brighter flares to have more negative values of $k$; the average value of $k$ for the 20 brightest flares is $-0.16$.

\subsection{Spectral variations}

In order to study variations of the photon index in detail, in Figure \ref{fig_scat_top20}, I show scatter plots of the normalised flux density $F_{\rm peak}$ vs. the photon index $\Gamma$ for each flare in the sample. Varying point intensity is used to show the flow of time. It is clear that individual flares are characterised by different average values of the photon index, and also by different spreads of its values. Scatter plots for long flares are obviously more complex than those for short flares. In some cases, the spectral evolution of the gamma-ray flare takes form of a clear loop (e.g. flare \#11 --- PKS~1510$-$089 at MJD~55868). Such loops were observed numerously in blazars in gamma rays \citep[\eg,][]{2010ApJ...721.1383A} and X-rays \citep[\eg,][]{2000ApJ...541..166F}. The most basic characteristics of a loop is its orientation in time. In Figure \ref{fig_scat_top20}, clockwise (CW) loops correspond to spectral hardening (decreasing $\Gamma$) at the peak, and counterclockwise (CCW) loops correspond to spectral softening (increasing $\Gamma$). I can identify clear cases of CW loops (flares \#8, \#9, \#11) and CCW loops (flares \#23, \#26, \#30). However, most cases are difficult to classify, and thus I cannot draw a definite conclusion about the statistics of loop orientations.

\subsubsection{Average photon index}

\begin{figure}
\includegraphics[width=\columnwidth]{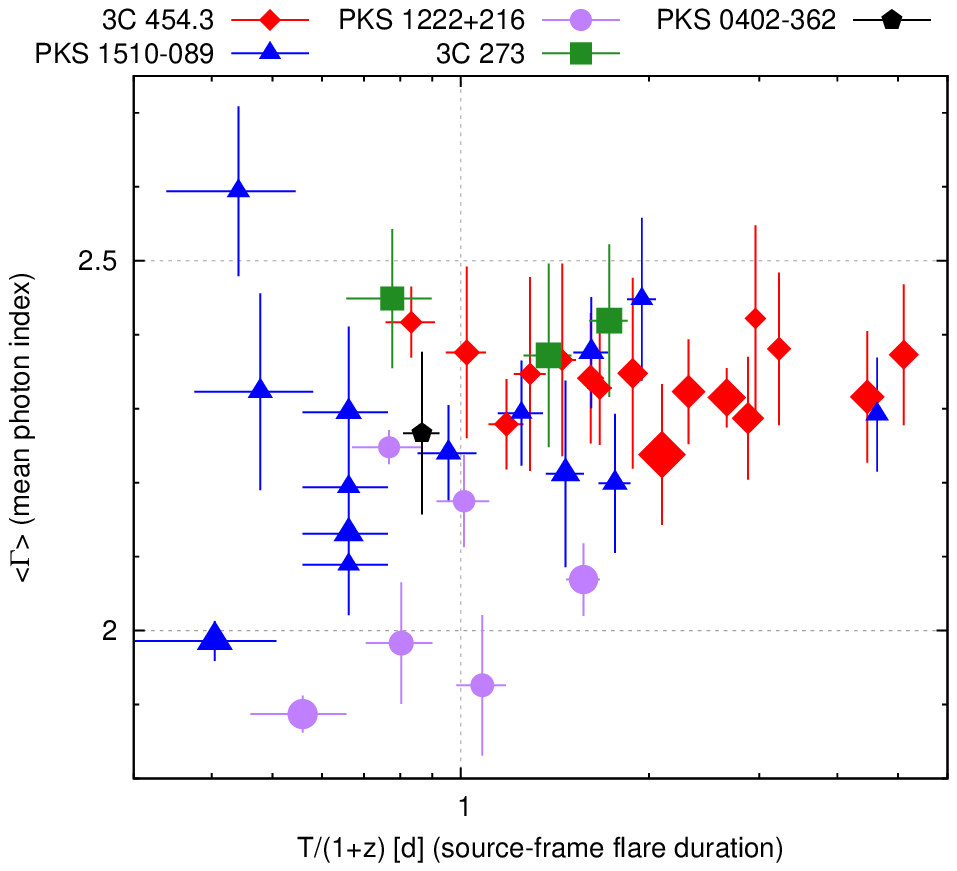}
\caption{Distribution of the source-frame flare duration $T/(1+z)$ vs. the average photon index $\left<\Gamma\right>$. The shape and colour of each point indicates the host blazar, and the point size is proportional to the logarithm of peak flux density $F_{\rm peak}$.}
\label{fig_scat_dur_ind}
\end{figure}

In Figure \ref{fig_scat_dur_ind}, I plot the distribution of the source-frame flare duration $T/(1+z)$ vs. average photon index $\left<\Gamma\right>$. The average photon index takes values between 1.9 and 2.6. The extreme values are realised only by very short flares, with $(1+z)T\lesssim 1\;{\rm d}$, and for long flares the range of $\left<\Gamma\right>$ converges to the vicinity of 2.3 -- 2.4. The average value of $\left<\Gamma\right>$ for the sample is $2.27$, and the median is $2.3$. Individual blazars produce flares of significantly different distributions of $\left<\Gamma\right>$. 3C~454.3 produces flares characterised by $2.2<\left<\Gamma\right><2.45$, i.e. with a hardness typical for the whole sample and with a very small scatter. PKS~1510$-$089 produces flares with a wide range of average photon index, $2\lesssim\left<\Gamma\right>\lesssim 2.6$. PKS~1222+216 produces relatively hard flares, with $1.9\lesssim\left<\Gamma\right>\lesssim 2.25$, and 3C~273 produces relatively soft flares, with $\left<\Gamma\right>\sim 2.4$. There is no clear dependence between the peak flux density $F_{\rm peak}$ and the average photon index.

\subsubsection{Photon index scatter}

\begin{figure}
\includegraphics[width=\columnwidth]{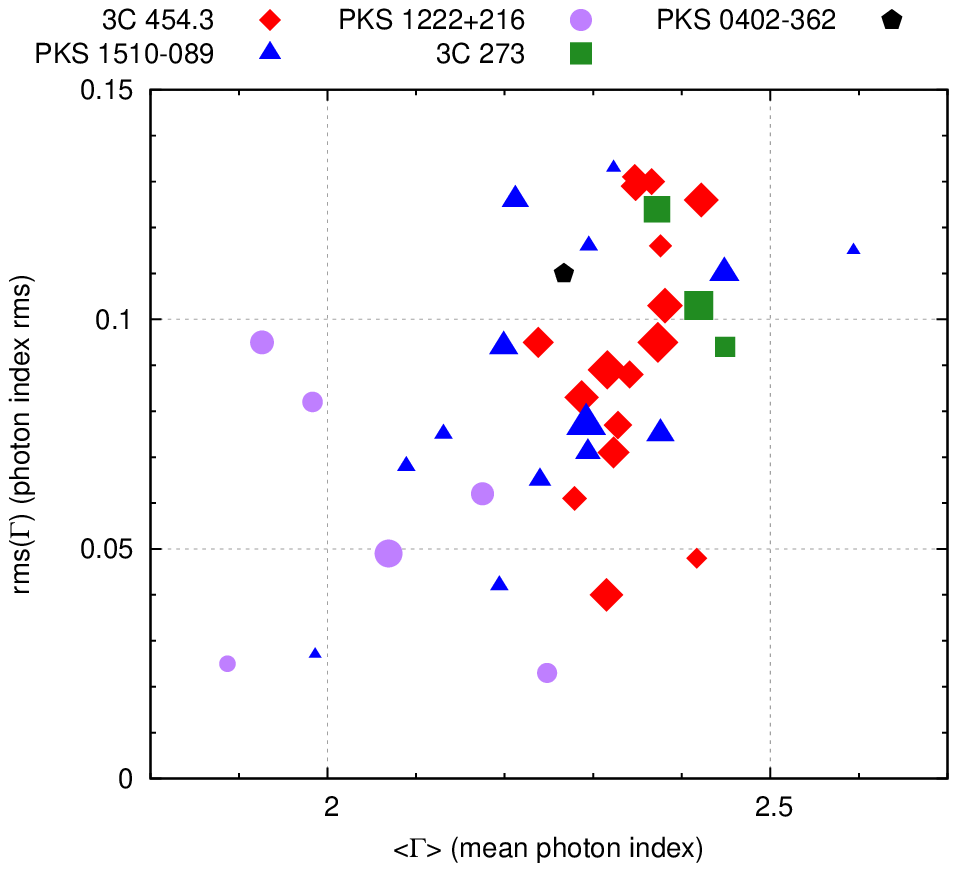}
\caption{Distribution of the average photon index $\left<\Gamma\right>$ vs. the photon index scatter ${\rm rms}(\Gamma)$. The shape and colour of each point indicates the host blazar, and the point size is proportional to the logarithm of source-frame flare duration $T/(1+z)$.}
\label{fig_scat_ind_rms}
\end{figure}

In Figure \ref{fig_scat_ind_rms}, I plot the distribution of the average photon index $\left<\Gamma\right>$ vs. the photon index scatter ${\rm rms}(\Gamma)$. The maximum value of the photon index scatter is $\sim 0.13$, and the minimum value is $\sim 0.02$. The average and median values are $\sim 0.09$. I find the general trend that the photon index scatter is higher for softer flares (higher $\left<\Gamma\right>$), the Pearson's correlation coefficient is $\rho(\left<\Gamma\right>,{\rm rms}(\Gamma)) = 0.5$. Individual blazars show a large range of ${\rm rms}(\Gamma)$ values. Flares of PKS~1222+216 are characterised by low photon index scatter (${\rm rms}(\Gamma) < 0.1$), while flares of 3C~273 show high average (${\rm rms}(\Gamma) > 0.09$). Flares of duration longer than $0.9\;{\rm d}$ in the source frame generally have ${\rm rms}(\Gamma) \gtrsim 0.05$, but the high photon index scatter values (${\rm rms}(\Gamma) > 0.1$) are exhibited by both long and short flares.

\section{Discussion}
\label{sec_dis}

Variability of blazars is a stochastic phenomenon, regardless of in which band and at what time scales it is studied. Even the very notion of a flare, while intuitive, is controversial, because it is difficult to formulate a simple and practical definition. Here, I propose a simple mathematical definition of a flare, based on the peak and half-peak flux levels, without considering the complexity of the observed light curves. Such defined flare may be fairly simple, with a single roughly time symmetric peak, or complex, with several local maxima and minima. In practice, even such a simple definition of a flare faces some problems: the flare duration and peak flux depend on the sampling (in the case of \emph{Fermi}/LAT data, the time binning), and limited photon statistics introduces some ambiguity in determining the flare beginning and end. Nevertheless, using the same definition for a uniform sample of blazar flares reveals new information about the variability of blazars.

The focus of this study is on individual flares hosted by a handful of blazars. Flares of any given blazar, separated by months or years, can be considered independent phenomena, as they are produced in constantly flowing relativistic jets. Any similarities and differences between them reflect independent realisations of the underlying processes of energy dissipation and particle acceleration. Similarities and differences between flares produced in different blazars reflect the environmental factors that vary over time scales of many years.

\subsection{Temporal properties}

The most appealing finding of this study is the systematic difference between flares produced by 3C~454.3 and other bright blazars. Flares produced by 3C~454.3 are significantly longer and more complex. A brief look at the first 10 panels of Figure \ref{fig_lc_top20} shows that they are characterised by multiple secondary maxima and minima. In accordance with the adopted definition of the flare, these minima are always higher than half of the peak flux for a given flare. The light curve of 3C~454.3 can be contained in a limited range of flux values for several days. Other blazars can produce such long flares only occasionally (e.g., flare \#28 in PKS~1510-089). The long flares of 3C~454.3 can be described as a collection of closely-spaced simple flares of comparable peak flux levels. On the other hand, short flares displayed mostly by other blazars appear to be simple (3C~454.3 can also produce simple flares, e.g. flare \#31).

\begin{figure}
\includegraphics[width=\columnwidth]{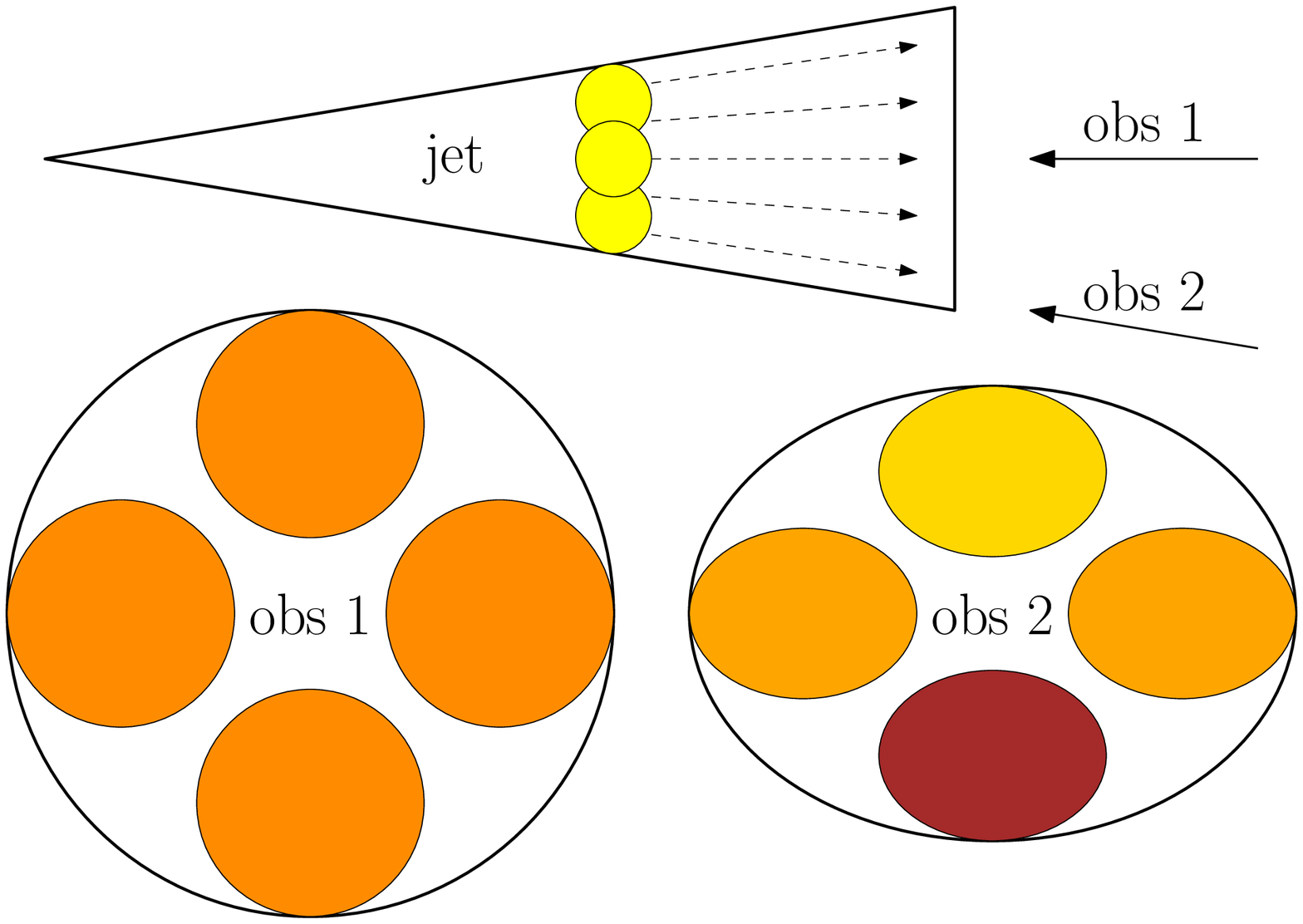}
\caption{Effects of the viewing angle $\theta_{\rm obs}$ on the relative Doppler boost (darker colour -- higher Doppler factor $\mathcal{D}$) of local emitting regions in a conically expanding relativistic jet. Observer 1 is located at $\theta_{\rm obs}\ll\theta_{\rm j}$, and observes similar Doppler factors from emitting regions located at different azimuthal angles. Observer 2 is located at $\theta_{\rm obs}\sim\theta_{\rm j}$ ($\sim\Gamma_{\rm j}^{-1}$) and observes different Doppler factors, and radiation from a more strongly boosted emitting region is observed earlier because of shorter light travel time. I propose that Observer 1 corresponds to us watching 3C~454.3, and Observer 2 corresponds to us watching other blazars producing the sample of bright flares.}
\label{fig_viewing_angle}
\end{figure}

The distinct structure of flares produced systematically by 3C~454.3 could be explained by its special jet orientation. \cite{2009A&A...494..527H} provide estimates of the jet Lorentz factor $\Gamma_{\rm j}=(1-\beta_{\rm j}^2)^{-1/2}$, where $\beta_{\rm j}=v_{\rm j}/c$, and viewing angle $\theta_{\rm obs}$ for a large sample of blazars monitored in the MOJAVE program. They estimate the viewing angle of 3C~454.3 at $\theta_{\rm obs}\simeq 1.3^\circ\simeq 0.45\Gamma_{\rm j}^{-1}$, which is significantly smaller than the values for PKS~1510-089 ($3.4^\circ$), PKS~1222+216 ($5.1^\circ$) and 3C~273 ($3.3^\circ$). The corresponding Doppler factor for 3C~454.3, $\mathcal{D} = [\Gamma_{\rm j}(1-\beta_{\rm j}\cos\theta_{\rm obs})]^{-1}\simeq 33\simeq 1.7\Gamma_{\rm j}$, is the highest among the blazars producing the brightest gamma-ray flares. This explains why 3C~454.3, at $z=0.86$, being more distant than other blazars, is the brightest of them. In fact, the viewing angle of the inner jet of 3C~454.3 could be even smaller than that estimated from the VLBI studies, as these estimates by their nature are not very precise, and the jets may be slightly bent even at parsec scales \citep{2002AJ....123.1258H}. The observed emission from a jet element propagating at the Doppler factor $\mathcal{D}$ is boosted by factor $\mathcal{D}^4$. For $\theta_{\rm obs}=0$ we have $\mathcal{D}=(1+\beta_{\rm j})\Gamma_{\rm j}$, while for $\theta_{\rm obs}=\Gamma_{\rm j}^{-1}$ (Doppler cone) we have $\mathcal{D}=\Gamma_{\rm j}$. The difference in the Doppler factors by factor $(1+\beta_{\rm j})\simeq 2$ corresponds to a difference in the observed fluxes by factor $\simeq 16$. For an average blazar, one can expect that $\theta_{\rm obs}\sim\theta_{\rm j}\sim\Gamma_{\rm j}^{-1}$, where $\theta_{\rm j}$ is the jet half-opening angle, so that the observer is located at random orientation within the Doppler cone\footnote{\cite{2009A&A...507L..33P} estimated the jet opening angles for the MOJAVE blazars. Their results indicate that $\Gamma_{\rm j}\theta_{\rm j}\ll 1$, which would not allow for high contrasts in the Doppler factors. However, this measurement, made at the frequency of $15\;{\rm GHz}$, probes jet regions located significantly beyond the gamma-ray emission sites, as indicated by a systematic delay of the $15\;{\rm GHz}$ radio signals with respect to the gamma rays \citep{2010ApJ...722L...7P}.}. In such a case, only a fraction of the jet volume is strongly Doppler-boosted and effectively contributes to the observed emission (see Figure \ref{fig_viewing_angle}). However, when $\theta_{\rm obs}\ll\theta_{\rm j}$, most of the jet volume is strongly Doppler-boosted. If individual simple flares are produced in different parts of the jet cross-section, their Doppler factors will be very similar, resulting in similar peak fluxes. Thus, the long complex gamma-ray flares of 3C~454.3 are consistent with its very close orientation to the line of sight.

Another interesting finding is that short flares ($T\lesssim 2\;{\rm d}$) tend to be time-asymmetric, with the flare onset shorter than the flare decay. In some cases, it can be seen that strongly asymmetric flares consist of a main peak and a subsequent secondary maximum. In the case of flare \#12 in PKS~1222+216, a hierarchical structure of simple flares is present. Why should secondary maxima follow the main peak, rather than preceding them? It can be understood in the scenario discussed in the previous paragraph. Emitting regions located at the same distance measured along the jet, but oriented at smaller angles to the line of sight, are located closer to the observer, so the light travel time is shorter. At the same time, they are more strongly Doppler boosted, therefore one observes a major peak followed by minor peaks. In the case of 3C~454.3, where $\theta_{\rm obs}\ll\theta_{\rm j}$, the Doppler factors of different emitting regions are similar, and thus its flares appear to be time-symmetric.

If the different jet orientations between 3C~454.3 and other blazars explain the major differences between parameters of flares produced by these sources, I can expect that blazar flares are intrinsically similar in all sources. The scenario presented in Figure \ref{fig_viewing_angle} assumes that the gamma-ray emitting regions in blazar jets are structured and consist of several domains. The long multiply-peaked flares of 3C~454.3 indicate that many simple flares of similar peak fluxes can be produced within a short range of time, so that they partially blend with each other. This partial blending may indicate that these simple flares are not entirely independent. A possible scenario for their origin is a large-scale jet perturbation that is intrinsically unstable or propagates through an inhomogeneous medium. Many major blazar flares have been associated with disturbances propagating along the jet that later-on are identified with superluminal radio knots \citep{2012arXiv1204.6707M}. Such disturbances can be interpreted as shock waves if the jet is weakly magnetised \citep{1985ApJ...295..358L}, or reconnecting magnetic domain interfaces if the jet is highly magnetised \citep{1997ApJ...484..628L,2011MNRAS.413..333N}. As most theoretical models for the formation of relativistic jets predict that they are initially significantly magnetised \citep{1977MNRAS.179..433B,1982MNRAS.199..883B}, and that it is difficult to efficiently convert the magnetic energy flux into kinetic energy flux \citep{1994ApJ...426..269B}, scenarios based on magnetic reconnection appear to be more likely. Even if the magnetic domains cannot be easily reversed in AGN jets, reconnection could be triggered by short-wavelength current-driven instabilities \citep[CDI;][]{1998ApJ...493..291B,2012arXiv1208.0007N}. On the other hand, shock waves can be efficient in magnetised jets if the jet features highly evacuated gaps \citep{2012MNRAS.422..326K}.

Variability of bright \emph{Fermi}/LAT blazars during the first year of the \emph{Fermi} mission was studied comprehensively by \cite{2010ApJ...722..520A}. In particular, they studied individual flares of 10 sources of high variability index. They used light curves integrated using 3-day bins and fitted them with double-exponential flare templates allowing for arbitrary time asymmetry. They found that the average flare duration was $\left<T\right>\simeq 12\;{\rm d}$ and the average flare asymmetry $\left<k\right>\simeq 0.08$, with broad distributions of these parameters. While both parameters were calculated in a different way than in this work, the flares studied by \cite{2010ApJ...722..520A} are different events than the flares studied here. By using time bins of 3 days and looking at entire light curves, they picked many long gently varying components. In this work, I use 0.5-day time bins and look only at pieces of the light curves that belong to individual flares according to my definition, and I find no flares longer than 10 days. While \cite{2010ApJ...722..520A} conclude that no particular type of flare asymmetry (positive/negative) is being favoured by their results, I show that only flares shorter than $\sim 1.5\;{\rm d}$ tend to be asymmetric.

\subsection{Spectral properties}

Average gamma-ray spectral index of the brightest blazar flares ranges between 1.9 and 2.6. As all of the studied blazars belong to the subclass of flat-spectrum radio quasars (FSRQs), these values are rather typical \citep{2011ApJ...743..171A}. I note that for the longest flares spectral indices converge into a narrow range between 2.3 and 2.4. This is most likely a statistical effect, as the average is calculated from a larger number of data points. Long flares show complex, chaotic structures in the gamma-ray flux -- photon index space, and the scatter of photon index values is of order $\pm 0.1$. But even short flares, while they usually have simple light curves, often show a substantial photon index scatter. Individual blazars show significant differences in the spectral behaviour. PKS~1222+216 typically shows relatively hard spectra with small photon index scatter, while 3C~273 shows relatively soft spectra with high photon index scatter. Overall, the gamma-ray spectral variability of flaring blazars is completely irregular. Because of lack of consistent patterns (e.g. CW or CCW loops), the flux -- photon index diagrams are not very useful as probes of the physics of particle acceleration in FSRQs.

\section{Conclusions}
\label{sec_con}

I selected a sample of the 40 brightest gamma-ray flares of blazars that occurred during the first 4 years of the \emph{Fermi} mission. A flare is defined as a period of time over which the flux exceeds half of the peak value. For each flare, I calculate its peak flux, peak luminosity, duration, time asymmetry, average photon index and the photon index scatter. I present light curves and flux--photon index diagrams, and study the distributions of the flare parameters.

The sample is hosted by 5 blazars, of which the most active were 3C~454.3 (16 flares) and PKS 1510-089 (14 flares). In general, flares produced by 3C~454.3 are longer and more complex than those produced by other sources. This can be attributed to a very small viewing angle of 3C~454.3 ($\theta_{\rm obs}\ll\theta_{\rm j}$), a consequence of which is that a larger fraction of the jet volume is strongly Doppler-boosted. Other blazars, having $\theta_{\rm obs}\sim\theta_{\rm j}$, produce shorter flares that are time-asymmetric, with the flare onset shorter than the flare decay. This is because jet regions that are more Doppler-boosted are located closer to the observer, and thus brighter subflares precede fainter ones. Long multiply-peaked flares indicate fragmentation of the gamma-ray emitting regions in blazar jets. I am unable to identify any regularities in the spectral variations of flaring blazars.

\section*{Acknowledgements}

I thank Aneta Siemiginowska, Mitch Begelman, Greg Madejski and Beno{\^i}t Cerutti for discussions, and the anonymous reviewer for helpful comments. This work was partly supported by the NSF grant AST-0907872, the NASA ATP grant NNX09AG02G, the NASA Fermi GI programme, and the Polish NCN grant DEC-2011/01/B/ST9/04845.

\end{document}